\documentstyle[preprint,aps,epsfig]{revtex}
\begin{document}
\tighten

\title{Localization of Quasiparticles in an NS Structure}
\author{ A.~V.~Shytov, P.~A.~Lee, and L.~S.~Levitov}
\address{
 12-112, Massachusetts Institute of Technology, Cambridge, MA 02139, USA
}
\maketitle
 
  \begin{abstract}
We discuss localization of quasiparticles in an extended NS
structure in the situation when the reflection from the NS
interface is mostly of Andreev kind. The localization of
quasiparticle states arises due to trajectory retracing caused by Andreev
reflection. This effect is semiclassical in the sense that in
the classical limit $\hbar\to0$ the states become fully
localized, while quantum diffraction effects {\it destroy} the localization
behavior. We derive the criteria for this localization and show
that it takes place only for sufficiently smooth disorder with
large correlation length, whereas short-range disorder destroys
the effect. Localization of quasiparticle states gives rise to
resonance peaks in the local tunneling density of states.
  \end{abstract}



We consider excitations in a thin normal metal film on the
surface of a superconductor. We assume that electrons with
energy well below superconducting gap $\Delta$ are reflected
from the NS interface by Andreev mechanism, which is the case if
the NS interface is clean and smooth (see Fig.~\ref{Fig1}). If
Andreev reflection occures {\it exactly}
backwards\cite{Andreev-1},
the electron orbit will be
closed. Such an electron will bounce between metal--vacuum and
metal--superconductor interfaces, after each two reflections
returning exactly to its original position. In such a model, all
trajectories appear to be localized. This simple fact can be
interpreted as integrability of classical Andreev
billiards\cite{chaos}.

Below we discuss quantum localization in this system. We add
details to the oversimplified picture discussed above and
correct it. It turns out that in the classical Andreev dynamics
localization (and integrability) is not a universal behavior. We
discuss complications arising in  the quantum problem, such as
the role of smooth disorder versus short--range disorder. We
compare the situation with that in a perfectly uniform metal
film, where the localization is absent. However, in presence of
smooth disorder with large correlation length localization can
take place. We derive the criterion for localization of this
type and discuss robustness of this phenomenon. 

Recently, the effect of Andreev reflection was discussed in the context of 
the problem of quantum chaos in NS cavities\cite{Brouwer,Bruder}. 
It was found that in the presence of Andreev scattering, 
chaotic dynamics gives rise to an energy gap centered at  
the Fermi level. This gap can serve as a benchmark of chaotic dynamics.
In contrast with the cavity problem, we study an infinite system.
In this case, a natural equivalent of ergodic (chaotic) dynamics in cavities
will be delocalization of states over the entire system. 
Accordingly, in the infinite system, 
an equivalent of regular dynamics in cavities is 
localization of states. 

We begin the discussion by reviewing properties of the states 
in a uniform film of constant thickness. In this system electronic
states are plane waves and thus are not localized. To see that,
one can solve Bogoliubov--de Gennes equations\cite{deGennes} for
this problem and derive the spectrum of
electrons\cite{Andreev-1}:
  \begin{equation}
\label{spectrum}
\epsilon_{n}(p_x) =
\frac{\pi \left(n + \frac{1}{2}\right) v_F}{L} \,
\sqrt{1 - \frac{p_x^2}{p_F^2}}\ .
   \end{equation}
Here $n$ is the quantum number ($n =0,1,2,\dots$),
$L$ is the thickness of the film,
$p_x$  is the momentum of the particle parallel to NS interface,
$p_F$ is Fermi momentum,
$v_F$ is Fermi velocity (here we set $\hbar = 1$).
The spectrum (\ref{spectrum})
has dispersion ($v(\epsilon)=\partial \epsilon/\partial p_x \neq 0$),
and therefore the states are not localized. However,
the dispersion is much weaker than for free electrons and thus
the Andreev states are much closer to becoming localized.

One can qualitatively understand this dispersion as follows.
When Andreev reflection occurs, an electron is converted into a
hole, and its energy (measured from $\epsilon_F$) changes sign.
Therefore, the momenta of electron and hole, $p_e$ and $p_h$,
are related by
  \begin{equation}
\label{momenta}
\frac{p_e^2}{2m} = \epsilon_F + \epsilon \ ; \qquad
\frac{p_h^2}{2m} = \epsilon_F - \epsilon
  \end{equation}
Here $\epsilon_F$ is Fermi energy, and $m$ is effective mass.
From (\ref{momenta}) it is clear that for the energy not right at
the Fermi level, i.~e. for $\epsilon \neq 0$, the momenta $p_e$
and $p_h$ are different. On the other hand, the component of
momentum parallel to the interface is conserved at Andreev
reflection. Consequently, the reflection angle must change. For
$\epsilon \ll \epsilon_F$ the change in the angle is small, and the
Andreev reflection law reads:
\begin{equation}
\label{snell}
\frac{\sin\theta_i}{\sin\theta_r} =
- \left(1 \pm \frac{\epsilon}{\epsilon_F} \right)
\end{equation}
($\theta_i$ and $\theta_r$ are the angles of incidence and
reflection measured from the normal to the interface).
Thus, after two subsequent reflections the particle will not
return to the starting point. Instead, it will be displaced
along the interface (see Fig.~\ref{Fig1}.a). The drift velocity
derived from this argument is the same as the one obtained above
from dispersion in (\ref{spectrum}).
Note that the same effect is responsible for the suppression of spectral 
flow in the superconducting vortex core\cite{Stone}.

Evidently, even though there is no localization in a uniform system,
by making the interface rough one can reach the situation
when classical Andreev trajectories will become localized, with
no drift along the interface. Thus our next step is to introduce
disorder in this model.

We consider here a model of disorder in which the thickness of
normal film is slowly varying. More precisely, we assume that
the normal metal--vacuum interface has some roughness, whereas
the NS interface is flat. The main effect of interface curvature
is that it acts like focusing mirror, counteracting the
dispersion. To overcome the dispersion, the curvature must
exceed certain threshold which will be estimated below.


Note that the localization effect we consider is based on the
semiclassical picture of a particle (nearly) retracing its
trajectory after being Andreev--reflected. Therefore, the
scattering by the surface roughness must also be semiclassical
in order to preserve the trajectory. The point is that quantum
effects in scattering, i.~e. diffraction, can destroy the
localization. Indeed, due to diffraction, quantum scattering is
stochastic and thus it violates reversibility of individual
trajectories. (Because two scattering events on the same
disorder configuration may not lead to identical results.) Thus,
we consider only sufficiently smooth surface fluctuations and
formulate below a quantitative condition on the degree of smoothness.

Suppose that the mean thickness of the normal metal film is
$L$, the variation of the thickness is $\Delta$, and the spatial scale
on which the thickness varies is $r_c$ (see Fig.~\ref{Fig2}).
The criterion for localization can be expressed in terms of these
parameters.

To begin with, let us ignore diffraction, and consider a purely
classical motion. Without any loss of generality, we can limit
the discussion to the problem in a two-dimensional space. We
will use coordinate system in which the NS interface is the line
$y =0$, and the metal--vacuum interface is given by $y = L(x)$.
It is very instructive to consider a metal--vacuum surface of
constant curvature, i.~e. of a spherical shape.

First, suppose
that the NS interface is in the equator plane. Consider a
trajectory which hits the NS interface exactly at the center of
the sphere (see Fig.~\ref{Fig3}). Note that any such trajectory
retraces itself
even if Andreev scattering does not occur exactly backwards. Thus,
for this trajectory the drift is indeed eliminated by the
curvature. One can also show geometrically that if the center of
hemisphere is within the normal region, i.~e., $L>R$, there will be
no net drift (see below). Thus, the situation when the center of curvature lies
exactly on the NS interface, is critical.

The conventional approach to dynamical problems of this kind
involves mapping of Poincare section\cite{Gutzwiller} in the
phase space. For the $D=2$ problem the phase space is
four-dimensional, but the energy conservation reduces the
dimension to three. Hence, Poincare section is two-dimensional.
To construct it, we consider the points on the metal--vacuum
interface hit by electron (and disregard the points due to
holes). These points are characterized by their $x$-coordinates
$x_i$. Also, we characterize the momentum of electron by the
angle $\theta_i$ between the momentum direction 
and the normal to the
surface at the collision point. Thus, each collision is
gives  a pair $(x_i, \theta_i)$,  and the trajectory of
the particle is represented as a sequence of points in the $(x,
\theta)$ plane. The Poincare section for a particular 
metal--vacuum surface shape is
shown in Fig.~\ref{Fig4}. This section exhibits typical
Kolmogorov-Arnold-Mozer features: stable periodic islands
representing finite motion, i.~e., non--escaping trajectories,
and the regions around the islands representing escaping trajectories. The
islands correspond to localized states, whereas the outer 
regions correspond to delocalized states.

To find the region of stability of localized orbits, 
one should consider the stability of a self-retracing trajectory, 
like the one in Fig.~\ref{Fig3}. Suppose that 
the particle starts from metal-vacuum interface 
at a point $x$ and its momentum direction is characterized 
by the angle $\theta$ defined above. 
Suppose also that $\theta \ll 1$.    
Using the reflection law (\ref{snell}) and simple geometrical considerations, 
one can write down the linearized equations for the phase space 
coordinates $x'$ and $\theta'$, describing the state of quasiparticle 
after after reflection from the NS interface 
and returning to the metal-vacuum interface: 
\begin{equation}
 \left(
  \begin{array}{c}
    x' \\
    \theta'  
  \end{array}
 \right) 
 = \hat{M} (\alpha) \left( 
   \begin{array}{c}
      x \\ 
     \theta
   \end{array}  
 \right) \ , 
\end{equation}
where $\alpha = \epsilon/\epsilon_F \ll 1$, and  
\begin{equation}
\hat{M}(\alpha) = \left(
               \begin{array}{cc}
                 1 - \alpha \frac{L}{R}  & \alpha L \\
                   -\frac{\alpha}{R} \left( 1 - \frac{L}{R}\right) & 
                              -1 + \alpha \left(1 - \frac{L}{R}\right) \\
               \end{array}  
            \right) \ . 
\end{equation}   
After one Andreev reflection the sign of $\epsilon$ changes, 
because electron turns into hole. 
Since 
in the construction of Poincare section 
we are interested only in electrons and not in holes, 
the matrix describing the motion 
in the $(x, \theta)$ plane, is $\hat{M}(\alpha) \hat{M}(-\alpha)$. 
In the first order in  $\alpha$ its eigenvalues are 
\begin{equation}
\lambda_{1,2} = 1 \pm 2 i \alpha \sqrt{1 - \frac{L}{R}} \ . 
\end{equation}
To  the same accuracy in $\alpha$, 
it is more natural to write 
\begin{equation}
\lambda_{1,2} = \exp \left( \pm 2 i \alpha \sqrt{1 - {L}/{R}}\right)
\end{equation}
to assure the phase volume conservation. 
The eigenvalues $\lambda_{1}$ and $\lambda_{2}$ 
are complex for $L < R$ and real otherwise. 
It means that the self-retracing trajectory is stable 
for $L>R$. 

To derive this relation in a more intuitive way, note that
for each orbit 
the maximal value of $x$ corresponds to $\theta = 0$, 
i.~e., to normal reflection.
Attentive reader will notice that here 
we have the same situation as in the above
example with the hemisphere.
When $x$ is maximal, which corresponds to the 
trajectory turning point, 
the trajectory is almost self-retracing, like that in 
Fig.~\ref{Fig3}.
The center of curvature at the point $x$
lies on the NS interface. 
We conclude that
localized states exist if the center of curvature
is above the NS interface.

The previous discussion shows that 
 the localization criterion is $L>R$, where $R$ is the
curvature radius. By estimating $R\sim r_c^2/\Delta$, one
arrives at
  \begin{equation}
\label{criterion-1}
r_c < r_c^{*} \sim \sqrt{L\Delta} \  .
\end{equation}
We call this condition {\it classical criterion} of localization.

The criterion (\ref{criterion-1}) is not the only constraint on
$r_c$. In this system, localized and delocalized classical
trajectories can coexist at the same energy. (Indeed, for any
value of $\epsilon$ there are trajectories going along straight
lines parallel to the NS interface.) Hence, any perturbation
which mixes these two types of states can destroy localization.
In particular, due to finite size of surface fluctuations which
focus electrons, {\it quantum diffraction}
takes place. Efect of diffraction on quantum chaos in 
non-superconducting systems was 
considered in \cite{Aleiner}. 
In NS structures the diffraction changes 
the orientation of kinetic momentum in a
random fashion, which leads to spreading of the states over the
whole system. We discuss manifesation of this effect 
below.

For smooth disorder, one can derive the criterion of
delocalization via diffraction by using (\ref{spectrum}) and
employing adiabatic approximation. For that, we make $L$ in
(\ref{spectrum}) position--dependent and interpret the energy
(\ref{spectrum}) taken at $p_x=0$ as spatially dependent
potential energy. Also, we expand the square root in
(\ref{spectrum}) and replace $p_x$ by $-i\partial/\partial_x$.
This gives effective kinetic energy. Thus, one gets Hamiltonian:
  \begin{equation}
\label{H-eff}
\hat{\cal H}_{{\rm effective}} = \frac{\pi \left(n + \frac{1}{2}\right) v_F}{L}
\left\{
   -  \frac{1}{2 p_F^2} \frac{\partial^2}{\partial_x^2}
   -  \frac{\delta L(x)}{L}
\right\}\ .
  \end{equation}
Here $\delta L(x) = L(x) - L$ is the deviation of the film thickness from its mean value.
Since the particle is localized near the thickest place, where $L(x)$ is maximal,
one can write: $\delta L(x)  \approx \Delta(1- x^2/r_c^2)$.
Then one can estimate the width $d$ of the ground state wave function
by comparing (\ref{H-eff}) to the harmonic oscillator problem:
  \begin{equation}
\label{wf-size}
d \sim \left(\frac{\lambda_F^2 r_c^2 L}{\Delta}\right)^{1/4}\ .
  \end{equation}
Quantum effects do not destroy a localized state if its smearing
given by (\ref{wf-size}) is much less than the
potential well width, i.e., if  $ d \ll r_c$.
Thus, one arrives at another condition:
   \begin{equation}
\label{criterion-2}
r_c >  \sqrt{\frac{L \lambda_F^2}{\Delta}}
\sim \frac{\lambda_F}{\Delta}\,r_c^{*} \ ,
   \end{equation}
where $r_c^{*}$ is defined in (\ref{criterion-1}).
This is {\it quantum criterion} of localization.

The classical and quantum criteria (\ref{criterion-1}) and
(\ref{criterion-2}) determine when localization can take place. 
Note that (\ref{criterion-1}) and (\ref{criterion-2}) 
are compatible only when
$\Delta>\lambda_F$. This is expected, because if the thickness
fluctuations are less than $\lambda_F$, the disorder cannot
separate a group of states out of  the continuum 
and localize these states. (In the 
case $\Delta < \lambda_F$, 
there is no room for an extra wavelength between the NS and
metal--vacuum interfaces).

Can one satisfy (\ref{criterion-1}) and (\ref{criterion-2}) in
a real system? The film thickness $L$ must be larger than, or
of the order of the superconducting coherence length $\xi_0$.
Otherwise, due to the proximity gap induced in the normal layer
there will be no excitations with the energies of interest.
Besides, the NS interface must have width $\ge\xi_0$ in order
that the reflection is fully Andreev. Thus, for a superconductor
with $T_c\sim 10K$ one gets $L \sim 1000 \AA$. Since
$\lambda_F < \Delta$ is required, let us take $\Delta \simeq
10\AA$. Thus, the criteria (\ref{criterion-1}) and
(\ref{criterion-2}) give $10\AA < r_c < 100\AA$. This means
that the surface must have a certain degree of smoothness. At
present, it is difficult to say how realistic this condition
is. One can imagine a situation where all abrupt jumps of the
surface are screened by conducting electrons so that the
resulting potential is sufficiently smooth.

Note that the localization of the type described above is
quite different from the usual one. First, it occurs only if
the scattering is classical, in contrast with  the usual Aderson
localization which is due to quantum nature of scattering.
A manifestation of that is the {\it suppression} of
localization in this system by short range disorder.
Note, however, that at sufficiently high impurity
concentration the electrons are again localized,
now by Anderson mechanism. Thus, localization is reentrant
with respect to disorder strength. We think that
at high impurity concentration Andreev reflection
should enhance Anderson localization effect. However, this question 
certainly needs more attention. 

Secondly, in this system there is no mobility edge: the energies of 
the localized and delocalized states are not separated.
This apparently contradicts the standard ergodicity argument by 
Mott\cite{Mott}
about the absence of coexistence of localized and delocalized states
with the same energy. The reason that there is no ergodicity\cite{Ergodicity} 
in our problem
is that the disorder is smooth. Due to this smoothness, there appear
adiabatic barriers
dividing the phase space into domains 
with very different dynamical characteristics (see Fig.~\ref{Fig4}).

Finally, due to the presence
of the superconductor, the localization in the NS structure is less
sensitive to Coulomb interaction effects. In fact, in this problem
we deal with charged quasiparticles in a highly conducting 
and thus well--screening medium.
Usually, the appearence of localized states 
at the metal-insulator transition in disordered
systems is controlled by effects of Coulomb interaction.
The reason for the importance of the  interaction is that 
due to poor conductivity near the transition 
the screening of the interaction is very slow. 
In contrast, in the NS structure 
the interaction is screened by the superconductor.
Therefore, the screening is always fast,
 no matter how slow individual localized electrons are.

Finally, we discuss how localization of this unusual type
could be observed. Perhaps, it cannot reveal itself through the
conductivity, because the superconductor will always shunt
electrical conductivity of 
the normal film. Instead, one can measure the
local tunneling density
of states, which could be probed, for example, by an STM.

If the normal film is flat, the spectrum of electrons is described by
(\ref{spectrum}). The average
density of states corresponding to (\ref{spectrum})
has the well-known sawtooth structure.
This structure should be independent of   the
STM tip position.

If the localized states are present, they
will add spatially dependent features to the tunneling density of
states. Each localized state will give rise to a peak in the density
of states 
if the STM tip is close to the place where
the state is localized.
Thus, one can study spatial correlation of peaks (and other
features) in the local tunneling
density of states.
If the peaks are really due to localized states,
they should be spatially uncorrelated.
This measurement scheme is insensitive to the supercurrent.

To summarize, we
studied localization of quasiparticles in a normal metal film
boarded by Andreev mirror. This system exhibits a new type
of localization caused by self-retracing due to Andreev reflection.
We derived the criteria for this  localization,
and discussed its manifestation in the tunneling density of states.

We are very grateful to the organizers of the workshop
``Directions in Mesoscopic Physics'' at the 
Lorentz Center at Leiden University, where part of this work was done. 
The work of A.~S. and L.~L. is supported by the MRSEC Program of 
the National Science Foundation under award number DMR 94-00334.
P.~A.~L. acknowledges the support of NSF grant number DMR-9523361.

\begin{center}
\figure{
\epsfig{file=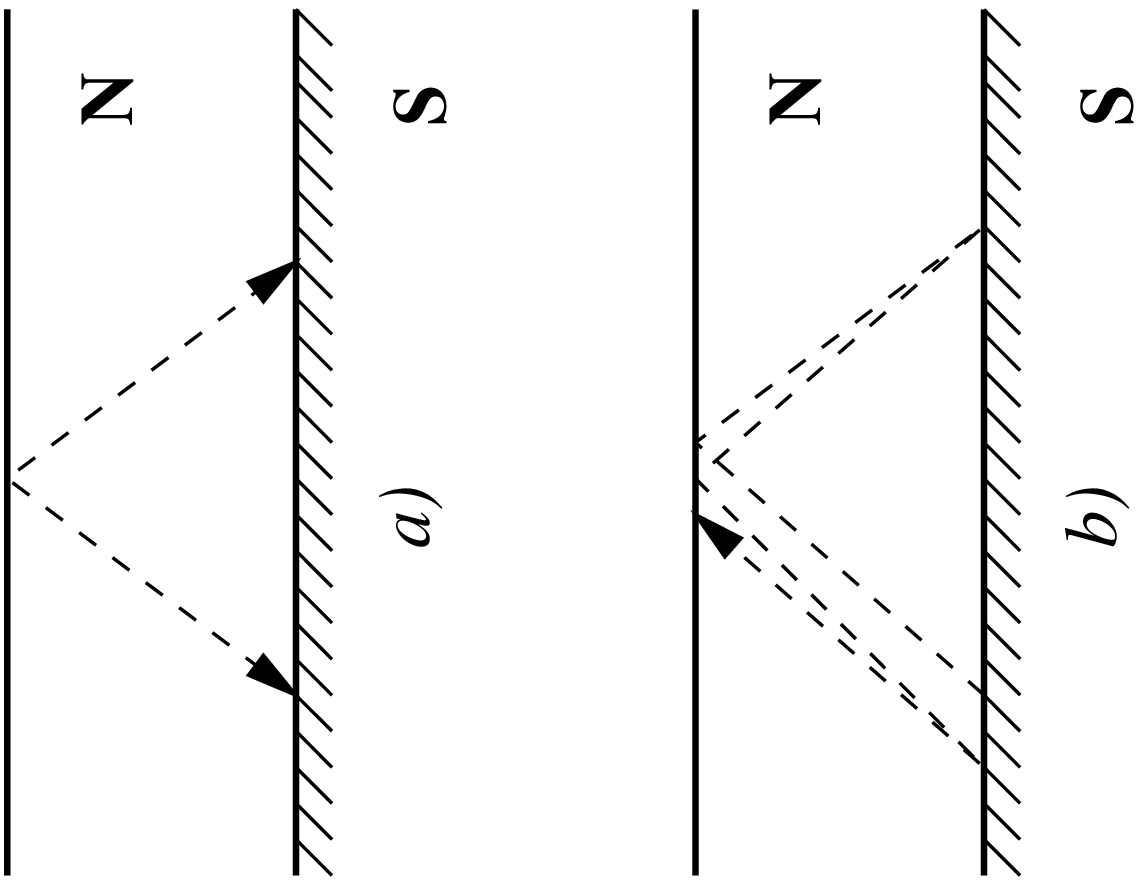, width = 150mm, angle=270}
\caption{(a) Idealized picture of localization.
The quasiparticle bounces between interfaces
without spreading over the whole system.
(b) At finite quasiparticle energy there is no perfect self-retracing
in Andreev scattering. This results in
a slow drift along the interface.}
\label{Fig1}
}
\end{center}

\begin{center}
\figure{
\epsfig{file=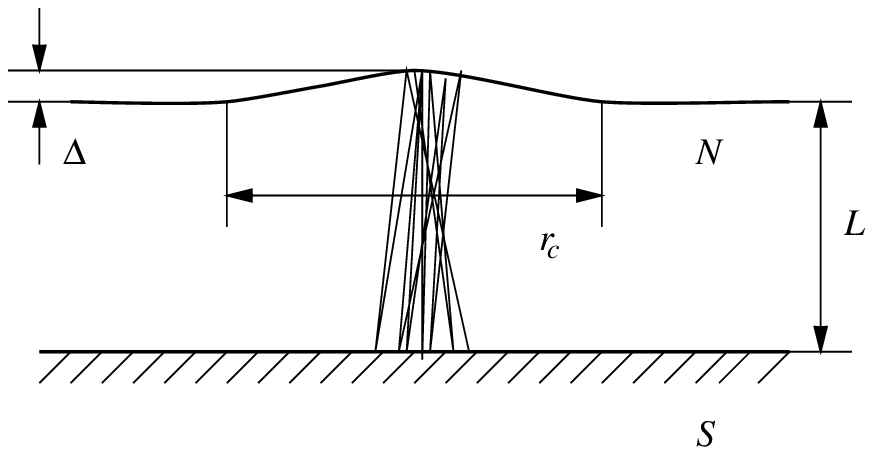, angle=270}
\caption{Localization of the quasiparticle trajectory
by surface fluctuation. The curvature of the
surface eliminates average drift.}
\label{Fig2}
}
\end{center}

\figure{
\epsfig{file=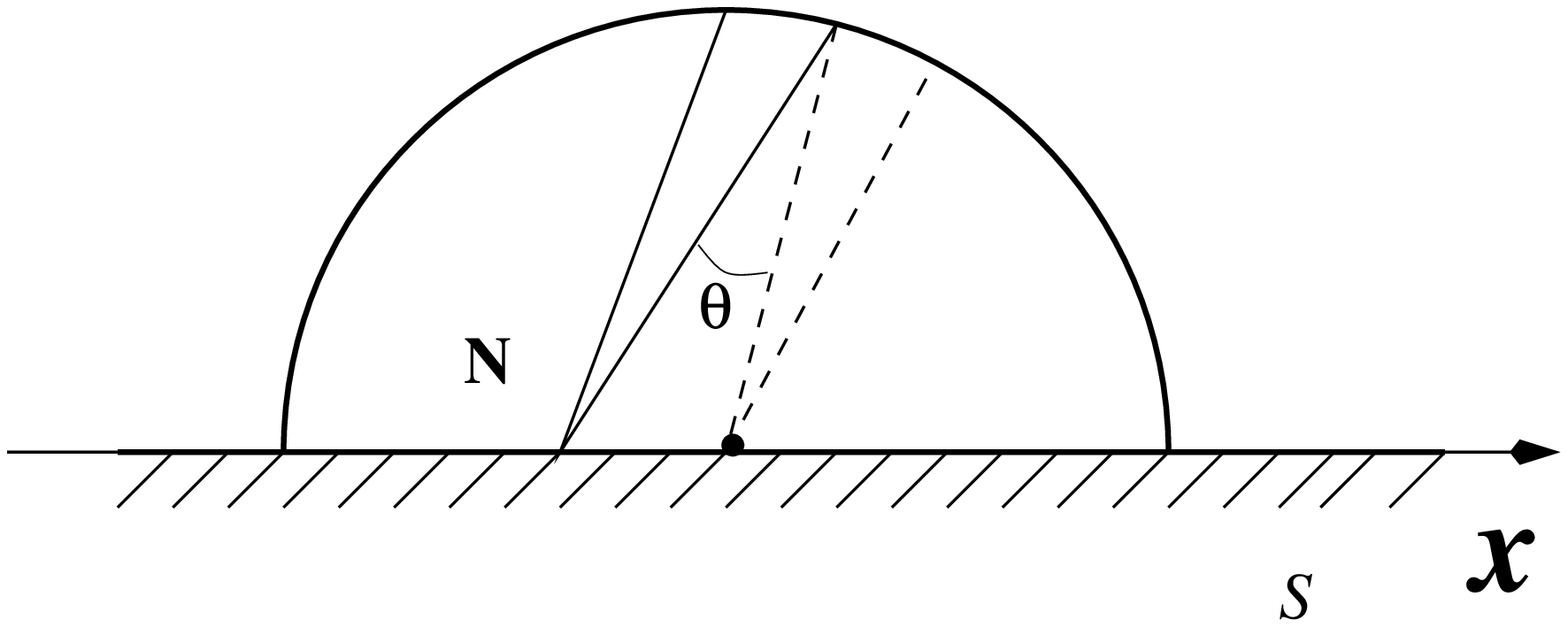, angle=270}
\caption{Particle bouncing inside a hemisphere.
When the center of curvature lies on the NS interface, 
any trajectory which passes through the center of curvature will
be self--retracing.}
\label{Fig3}
}

\figure{
\epsfig{file=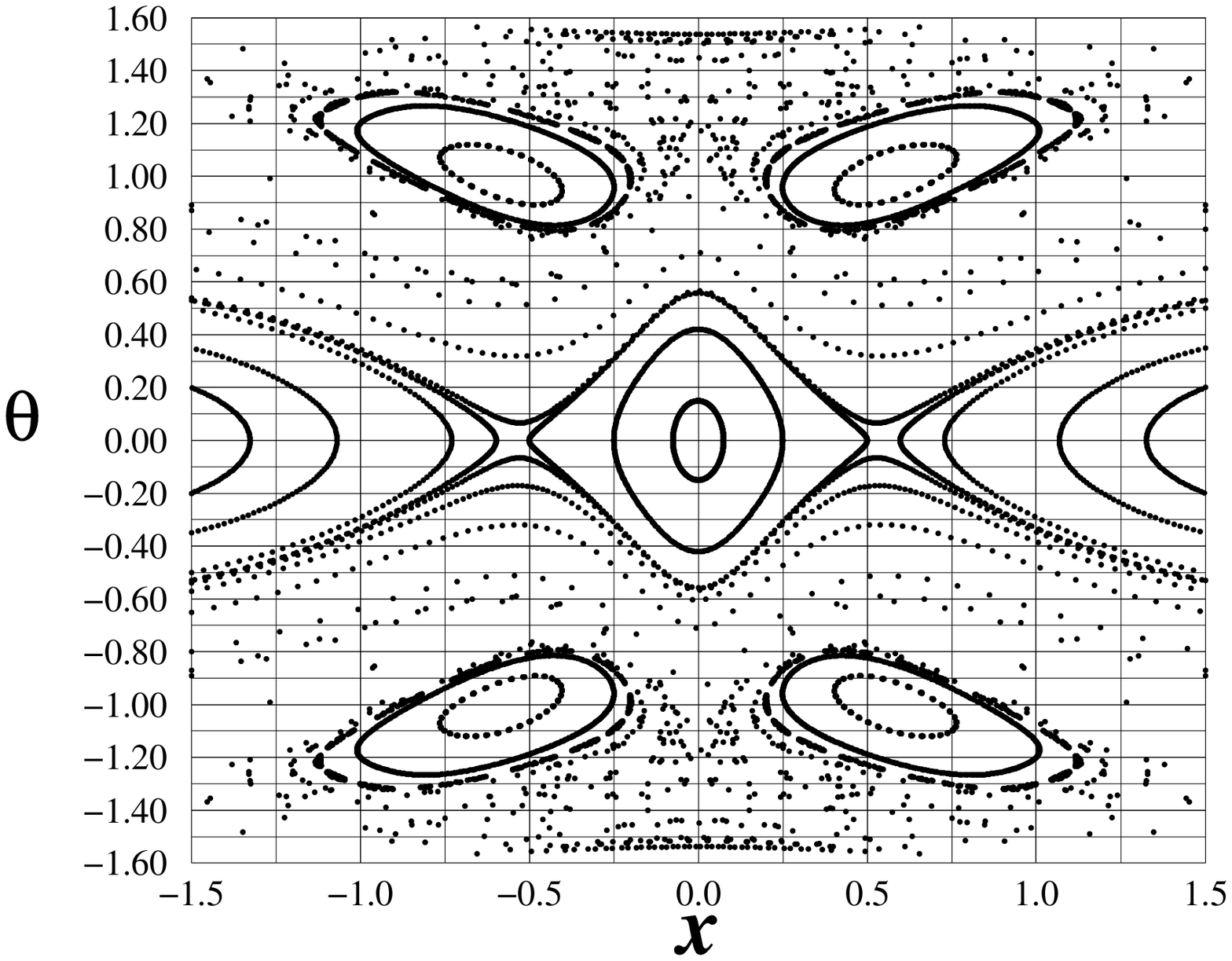, width=150mm,  height=150mm}
\caption{Poincare section for $y(x) = 10 + 1/(1+x^2)$, 
$\epsilon = 10^{-3} \epsilon_F$.
The central island 
and the four islands located symmetrically around it 
correspond to different periodic trajectories. 
If one denotes each collision with NS interface by S, 
and with metal--vacuum interface by V, then 
the central island corresponds to VSVSVS\dots \ sequence of collisions, 
while the lateral ones correspond to VVSVVS\dots \  sequence. 
}
\label{Fig4}
}


\begin{thebibliography}{99}
\bibitem{Ergodicity}
   In our problem the matrix elements between the states 
   localized near  different maxima of $L(x)$ are exponentially 
   small due to smoothness of the fluctuations of $L(x)$, i.e., 
   due to adiabaticity of the motion. 
   Still, since  these matrix elements 
   are non-zero, we do not have  true localization here, 
   and, strictly speaking, the Mott argument still holds.
   Thus, the states we discuss are only quasilocalized 
   with exponentially large decay times.
\end{thebibliography}
\end{document}